\documentclass[reprint,superscriptaddress,twocolumn, aps, prb]{revtex4-1}

\usepackage{graphicx,psfrag}
\usepackage{epstopdf}
\usepackage{amsmath} 
\usepackage{bm}
\usepackage{amssymb}
\usepackage{subfigure}
\usepackage{color}
\usepackage[pagewise]{lineno}
\usepackage{braket}% bra ket notation

\begin{document}

\title{\large \bf Design and Experimental Observation of Valley-Hall Edge States in Diatomic-Graphene-like Elastic Waveguides
}

\author{Hongfei Zhu}
\email{Hongfei.Zhu.44@nd.edu}
\affiliation{Department of Aerospace and Mechanical Engineering, University of Notre Dame, Notre Dame, IN 46556, USA}
\affiliation{Ray W. Herrick Laboratories, School of Mechanical Engineering, Purdue University, West Lafayette, Indiana 47907, USA}

\author {Ting-Wei Liu}
\email{liu2041@purdue.edu}
\affiliation{Ray W. Herrick Laboratories, School of Mechanical Engineering, Purdue University, West Lafayette, Indiana 47907, USA}

\author {Fabio Semperlotti}
\email{To whom correspondence should be addressed: fsemperl@purdue.edu}
\affiliation{Ray W. Herrick Laboratories, School of Mechanical Engineering, Purdue University, West Lafayette, Indiana 47907, USA}

\begin{abstract}
	
We report on the design and experimental validation of a two-dimensional phononic elastic waveguide exhibiting topological Valley-Hall edge states. The lattice structure of the waveguide is inspired by the diatomic graphene and it is imprinted in an initially flat plate by means of geometric indentations. The indentations are distributed according to a hexagonal lattice structure which guarantees the existence of Dirac dispersion at the boundary of the Brillouin zone. Starting from this basic material, domain walls capable of supporting edge states can be obtained by contrasting waveguides having broken space inversion symmetry (SIS) achieved by using local resonant elements. Our theoretical study shows that such material maps into the acoustic analog of the quantum valley Hall effect (QVHE) while numerical and experimental results confirm the existence of protected edge states traveling along the walls of topologically distinct domains. 

\end{abstract}

\maketitle

\section{Introduction}
In recent years, the study of topological phases of matter \cite{ReviewKaneTI,nobel} has inspired researchers across the most diverse fields of science and engineering. The possibility to design materials capable of achieving ideal levels of transmission even in presence of imperfections and defects would have a profound impact on many practical applications and open the way to the design of innovative devices. While this area of research originated in quantum physics, it recently expanded to include also the acoustic behavior of fluids and solids \cite{TopoPhon-Gyro, TopoGyroExp,TopoSound-Flow, TopologicalAcoustics-Flow, TunableTopoPnC-Flow,MechanicalTI, AcousticTIPlate, AcousticTIAir,ValleySonicBulk, ValleySonicEdge, Ruzz1, Ruzz2,TWarXiv}.

In solid state physics, one of the first implementations of topological materials was based on the use of the Quantum Hall Effect (QHE) which exploited an external magnetic field to break time-reversal symmetry (TRS). While in electronic systems the application of a magnetic field is a rather simple way to achieve TRS breaking, in acoustics the process is more complicated due to the intrinsic reciprocal nature of acoustic waves. The first few attempts at breaking TRS in an acoustical system exploited the use of rotating inclusions (such as spinning rotors \cite{TopoPhon-Gyro, TopoGyroExp} or fluids \cite{TopoSound-Flow, TopologicalAcoustics-Flow, TunableTopoPnC-Flow}). Later versions of topological acoustic systems were developed based on the acoustic analogue of the Quantum Spin Hall Effect (QSHE) \cite{MechanicalTI, AcousticTIPlate, AcousticTIAir} which, contrarily to the QHE, required intact TRS. In this latter case, researchers were able to achieve unidirectional edge states topologically protected from back-scattering by creating acoustic pseudo-spins and pseudo-spin-dependent effective fields.

More recently, several studies have investigated the design of topological materials based on broken space-inversion symmetry (SIS). The creation of such edge states cannot be explained by the previously mentioned QHE or QSHE mechanisms. In fact, when only SIS is broken the lattice still possesses a trivial topology within the context of QHE \cite{ReviewKaneTI, Raghu, PhotonicGraphene} and QSHE \cite{Z2, SpinChernHaldane}. However, due to the large separation in $k$-space of the two valleys (i.e. the Dirac points occurring at the corners of the Brillouin zone, the $\mathbf{K}$ and $\mathbf{K}'$ symmetry points), valley-dependent topological invariants can be defined and used to classify the topological states of the different lattices. This approach, usually referred to as quantum valley-Hall effect (QVHE), was recently investigated also for application to fluidic and elastic acoustic waveguides \cite{ValleySonicBulk, ValleySonicEdge, Ruzz1, Ruzz2, TWarXiv}.

While most of the above studies have concentrated on electromagnetics and acoustics, the theoretical and experimental observation of topologically protected edge states (TPES) in elastic media has been fairly limited due to the unique challenges occurring in these systems. Designs based on QHE need active components (e.g. gyroscopes or fluid circulators) or the application of an external field (e.g. magnetic field) to break time reversal symmetry. This approach makes the system very complex and often difficult to implement in practical applications. On the contrary, QSHE-based designs employ fully passive mechanisms but often results in non-trivial geometric and/or material configurations in order to engineer the band structure that requires a finely tuned double Dirac cone\cite{AcousticTIPlate}. The QVHE, however, relies only on breaking the space inversion symmetry, which is relative easier to achieve in elastic systems of practical interest. Only recently Vila et al. \cite{Ruzz1,Ruzz2} reported the first experimental realization of TPES based on QVHE in an elastic medium \cite{Ruzz2}. Their design exploited a hexagonal truss-like lattice having local masses attached on selected locations in order to break the space inversion symmetry. No further experimental studies are reported in the current literature concerning the use of QVHE for the design of topological elastic waveguides. In particular, while Vila's work [\onlinecite{Ruzz2}] proved the feasibility of the QVHE for elastic media, the design and experimental validation concerned only truss-like lattices that are not well-suited for structural applications. Recent studies \cite{Zhu,Semperlotti,Zhu2,ZhuPRB,ZhuPRL,ZhuJAP} have suggested that thin-walled metastructures can have important applications to the control of vibrations and structure-radiated sound in lightweight structures, such as those of interest for aerospace applications or, more in general, for high-performance transportation systems. In order to be able to extend topological acoustic designs to this class of applications, the structural waveguides will have to be fully continuous so to be able to sustain distributed loads such as those produced, for example, by aero- or hydrodynamic forces.

In the present study, we explore and experimentally validate the design of a fully-continuous and load-bearing phononic structural waveguide based on QVHE and capable of one-directional guided modes along the walls of topologically distinct domains. The fundamental design leverages geometric indentations distributed in a graphene-like arrangement within an otherwise flat thin plate. The resulting lattice structure has a $C_{3v}$ symmetry that fully preserves SIS. The symmetry of the lattice can be lowered to $C_3$ by introducing locally-resonant elements located at selected sites. This structure can be considered as an elastic analogue of a diatomic graphene, as shown in Fig. \ref{Fig1}a.
Depending on the location of the resonant elements, two configurations (later referred to as states) can be defined (Fig.\ref{Fig1}a and \ref{Fig1}b). In the following, we will show that these configurations (that are mirror images of each other) are topologically different and can be contrasted in order to create domain walls able to support topological modes. 

 We will study the characteristics and behavior of this material via a combination of theoretical, numerical, and experimental results to show that 1) the evolution from one state to the other is accompanied by a topological phase transition, and that 2) edge states are supported along domain walls (DW), that is those interfaces between adjacent phases.
In contrast with previous studies, we analyze the dynamic behavior of the QVHE waveguide by using a fully continuum modeling approach which provides a general methodology of analysis and allows mapping the topological behavior to the fundamental massless Dirac equation. We present also an in-depth study of the edge states and of their coupling with both the background material and the states supported by different domain walls. Finally, we experimentally study the propagation of topologically protected edge states in a nearly lossless elastic system and the effect of short range disorder. This analysis would complement and expand the results from previous studies on lossy elastic media \cite{Ruzz2} where the higher damping levels does not allow a definitive assessment of the disorder-induced back-scattering.

From a general perspective, we highlight that the proposed waveguide design fully preserves the structural integrity and the load-bearing capabilities of the host medium while, at the same time, it enables a high degree of flexibility in tailoring the topological properties. Such design will open up a broader spectrum of practical applications.

\section{Numerical Results}
\subsection{Band Structure Analysis}

\begin{figure}[h]
	\includegraphics[scale=0.44]{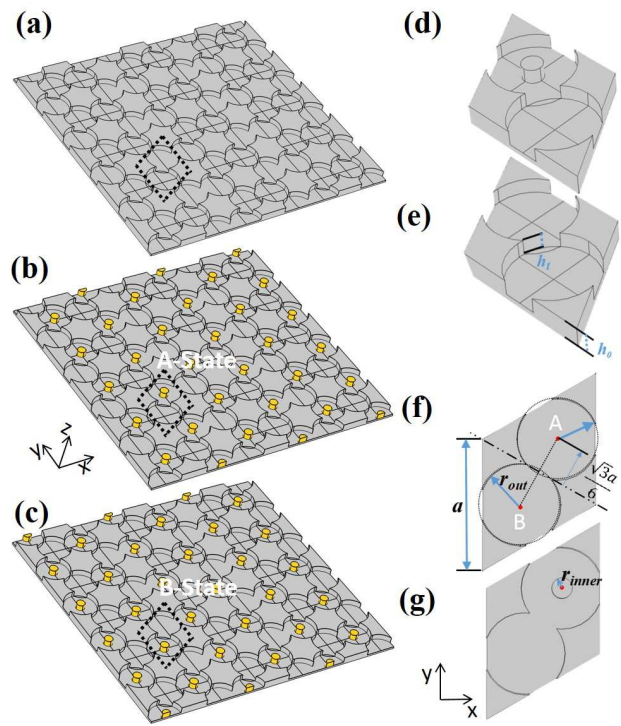}
	\caption{Schematic view of the fundamental unit cells and corresponding lattice structures. (a) shows the reference crystal structure which preserves space inversion symmetry. The lattice configuration is imprinted by engraving blind circular holes according to a graphene-like pattern on an initially flat aluminium plate. (b) and (c) show the two possible states achieved from the original lattice structure by breaking SIS. These configurations are the elastic analogue of a diatomic graphene lattice and are obtained by adding locally resonant elements at selected sites. These configurations are classified as $A-$ or $B-$states depending on the site ($A$ or $B$ as shown in (f)) the was modified. The black dashed box indicates the primitive unit cell in each of the three lattice structures. (d) and (e) show the isometric view of the unit cell of both the unperturbed and the $A-$state lattices. (f) and (g) show the top-view of the same unit cells and define basic geometric parameters.} \label{Fig1}
\end{figure}

The fundamental diatomic graphene-like unit cells and their corresponding lattice structures are shown in Fig.\ref{Fig1}. The unit cell has lattice constant $a=27.07$ mm, radius $r_{out}=8.46$ mm, and thickness inside the indentation $h_t=3.045$mm. The background medium, i.e. the plate, has thickness $h_0=4.06$ mm. The units were assumed to be made out of aluminum having mass density $\rho$ = 2700 $kg/m^3$, Young’s modulus $E$ = 70 GPa, and Poisson’s ratio $\nu$= 0.33. 

In order to break the SIS, we alter one of the indentation by adding a local mass in the form of a cylindrical pillar having radius $r_{inner}$. In other terms, the initial blind circular hole becomes a blind ring with the same outer radius and thickness. The pillar can be added either to site $A$ or $B$ (see Fig. \ref{Fig1}f), therefore forming two possible material states (see Fig. \ref{Fig1}b and \ref{Fig1}c) that have the same geometric configuration and are mirror images of each other.

\begin{figure}[h]
	\includegraphics[scale=0.43]{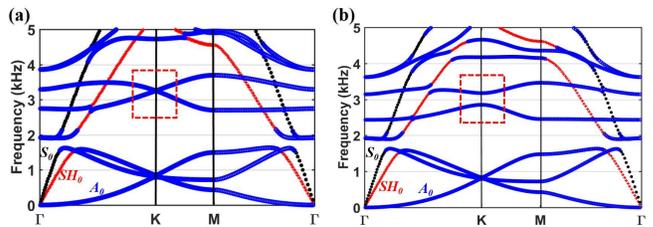}
	\caption{Phononic band structure of (a) the monoatomic graphene-like lattice (Fig.\ref{Fig1}(a)), and of (b) the diatomic graphene-like lattice (either $A-$ or $B-$state). Different mode types are identified by different colors, SH = shear horizontal mode (red), A= antisymmetric mode (blue), and S = symmetric mode (black). A complete bandgap of the fundamental flexural modes $A_0$ opens up at the original Dirac point when SIS is broken (see red dashed box).} \label{Fig2}
\end{figure}

The dispersion relations were calculated using a commercial finite element solver (Comsol Multiphysics). Given the finite dimension of the unit cell in the thickness direction the dispersion curves are composed by symmetric (S), anti-symmetric (A), and shear horizontal (SH) guided Lamb modes. The resulting dispersion curves are shown in Fig.\ref{Fig2} where different colors were used to identify the different wave types. Mode hybridization between $A$ and $S$ modes \cite{ZhuPRB} is clearly observed and it is due to the lack of symmetry of the unit cell with respect to the neutral plane. Although the presence of hybridization does not prevent achieving TPESs, we mention that its occurrence could be avoided by employing symmetric taper configurations with respect to the neutral plane. We specifically pursued a single-sided design for the taper because it yields more general configurations and illustrates how topological properties could be effectively obtained while maintaining a fully flat surface. This latter aspect has important implications in applications where the aerodynamic character of the thin panel must be preserved.

The analysis of the dispersion properties in Fig.\ref{Fig2}a reveals the existence of a degeneracy at the $\boldsymbol{K}$ point from which locally linear and isotropic dispersion curves emanate. In $k-$space the curves identify two cones that touch at their vertices at the frequency corresponding to the degeneracy. The cone-like dispersion structure is the Dirac Cone (DC) while the degeneracy is the Dirac Point (DP). This phenomenon is a direct consequence of the $C_{3v}$ lattice symmetry, therefore implying that the DC is protected by the lattice configuration. When the symmetry is lowered to $C_3$ by introducing a resonant pillar (either at site A or B), the SIS is broken and the DP degeneracy is lifted giving rise to a complete bandgap for $A_0$ modes. This mechanism is illustrated in Fig. \ref{Fig2}b that shows the band structure of the SIS-broken lattice (for either the A- or B-state). In these simulations, a radius $r_{inner}=2.12$ mm of the pillar was used. Note that the larger $r_{inner}$ the wider the topological bandgap that opens up at the degeneracy. It should be kept in mind that larger perturbations of the original geometry also result in stronger inter-valley mixing which ultimately deteriorates the immunity of the edge modes from backscattering. This specific aspect will be further addressed in later sections. 

\subsection{Berry curvature and valley Chern number}

The $A-$ and $B-$state lattices belong to different topological phases \cite{TWarXiv} and can be assembled together in a single lattice in order to enforce a topological phase transition along the wall separating them. In such case, when moving across the transition from one state to the other the topological bandgap would first vanish and then reopen again when entering the opposite state. The topological nature of this transition can be characterized using a topological invariant known as the Chern number $C_n$.

The parameter $C_n$ is obtained by integrating the Berry curvature $\Omega _n(\mathbf{k}) = \nabla _\mathbf{k} \times \braket{\mathbf{u}_n(\mathbf{k})|i\nabla _\mathbf{k}|\mathbf{u}_n(\mathbf{k})}\cdot \hat{\mathbf{z}}$ of the $n^\textrm{th}$ mode throughout the first Brillouin zone. For this system, $C_n$ is expected to be zero due to an odd distribution of the Berry curvature in $k$-space, which should be expected given that TRS is preserved \cite{ReviewKaneTI}. It follows that these lattices are classified as topologically trivial in the context of QHE systems. Nevertheless, for perturbative SIS breaking the Berry curvature peaks in correspondence to the valleys (i.e. around the $\boldsymbol{K}$ and $\boldsymbol{K'}$ symmetry points), having the form \cite{ValleyContrasting}

\begin{equation}\label{omega_b}
\Omega(\mathbf{q}) =\pm \frac{1}{2} m v_g^2 (\lvert \mathbf q\rvert^2 v_g^2+m^2)^{-\frac{3}{2}},
\end{equation} 

where $\mathbf{q}=\mathbf{k}-\mathbf{k}_{K/K'}$ is the wavevector deviation form the corresponding valley point, $v_g$ is the group velocity at the valley point, and $m$ indicates the strength of the SIS breaking (i.e. it is directly correlated to the size of the cylindrical pillar). The sign depends on the choice of the state, of the valley point, and of the mode emanating from the DC. The integral of the Berry curvature can be analytically calculated to be $\pm \pi$, as the bounds of the integral extend to infinity. This value is independent of the parameters $v_g$ and $m$. In such case, it is possible to define another topological invariant defined in the neighborhood of the valley that is known as the valley Chern number $C_v$. The $C_v$ of the $n^\textrm{th}$ band is defined as

\begin{equation}\label{Cv}
2\pi C_v=\int \Omega _n (\mathbf{k}) d^2 \mathbf{k}
\end{equation}
where the integral bounds extend only to a limited area around the valley. When the SIS breaking term $\lvert m\rvert$ is small, the Berry curvature is strongly localized around the valley and the integral converges quickly (i.e. in a small area around the valley) to a value $C_v=\pm 1/2$. This quantized value characterizes the bulk\textendash edge correspondence \cite{ControlEdgeStates} while the difference $\Delta C_v=C_{v,up}^A-C_{v,up}^B$ between the valley Chern numbers of the upper (or lower) bands of two adjacent lattices indicates the number of gapless edge states expected at the domain wall. This analysis suggests that an edge state should be expected at the DW between adjacent $A$- and $B$-state lattices, because their corresponding Berry curvatures result in $\Delta C_v=1$.

\begin{figure}[h]
	\includegraphics[scale=0.43]{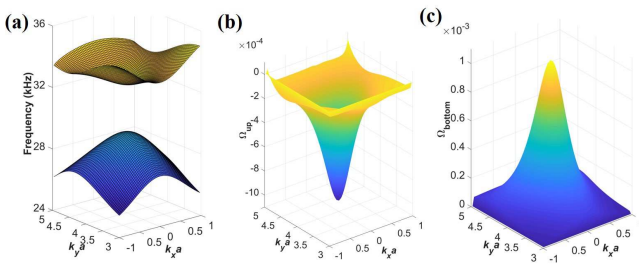}
	\caption{ (a) The dispersion surfaces of the flexural $A_0$ modes of the $A$-state lattice near the $\boldsymbol{K}$ valley. (b) and (c) show the Berry curvatures of the upper and lower $A_0$ modes once the degeneracy at the DP is lifted.} \label{Fig3}
\end{figure}

\begin{figure}[h]
	\includegraphics[scale=0.6]{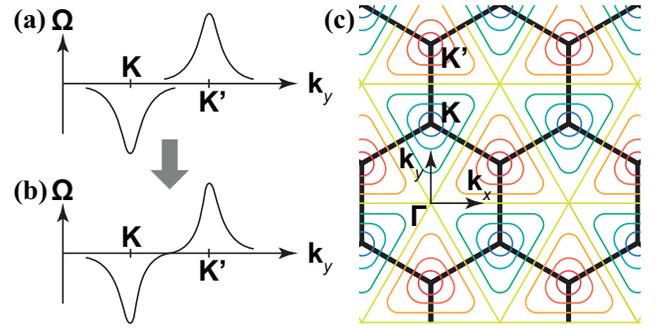}
	\caption{(a) The Berry curvature functions at two neighboring valleys in the case of large SIS-breaking. (b) The resulting Berry curvature (i.e. the sum of the contributions from both valleys) is zero around the inter-valley center. (c) The Berry curvature contours on the $k$-plane. The contours evolve from circular- to triangular-shaped while moving from the valley centers towards the bisectors of the valleys, where the Berry curvature vanishes.} \label{BCContour}
\end{figure}

Using Eqn. (\ref{omega_b}), we carried out the numerical calculation of the Berry curvature of the $A$-state lattice, for both the upper and lower mode. The calculation was performed over a finite squared area around the $\boldsymbol{K}$ point with $q_xa,q_ya\in (-1.67, 1.67)$, as shown in Fig.~\ref{Fig3}b, and the resulting Berry curvature was integrated to obtain the valley Chern numbers. The resulting valley Chern numbers are $\pm$ 0.35 for both the upper and lower mode, that is approximately 30\% error from the theoretical quantized value $C_v=$1/2. The reason for this discrepancy is found in the relatively strong SIS-breaking that was applied to the lattice. In the case of strong symmetry breaking, $\lvert m\rvert$ increases in Eqn. (\ref{omega_b}) which results in a broader (less localized) distribution of the Berry curvature function around the $\boldsymbol{K}$ point. It follows that, not only the integral of $\Omega_n (\mathbf{q})$ converges slower, but the Berry curvature extends across the inter-valley center ($\boldsymbol{M}$ point). This \textit{over-extension} of the Berry curvature cancels out the Berry curvature with opposite sign (associated with the neighboring valley) as shown in Fig.~\ref{BCContour}(a,b). In a 2D $k$-plane view, it is also observed that the Berry curvature contours evolve from circular- to triangular-shaped when moving from the valley centers towards the bisectors of the valleys, where the Berry curvature vanishes. This analysis clarifies why the calculated valley Chern number is always less than its expected quantized value and highlights that in presence of strong SIS-breaking the valley Chern number is not uniquely defined. Nevertheless, we note that the calculated value of $C_v$ still provides useful insights with regards to the characterization of the lattice structure. The strong SIS-breaking also leads to edge states that do not cross completely the topological band gap and that are subject to weak reflections when traveling along sharp corners (i.e. in conditions of high-disorder), as shown in later sections.

\subsection{Domain-wall edge states}

The analysis presented above indicates that by connecting $A$- and $B$-state lattices along selected edges of the graphene-like lattice (i.e., zigzag, armchair, bearded) topologically protected edge states should be expected along the DW interface. In the following, we present the case of domain walls assembled from zigzag edges as an example of possible topological lattices obtainable from the fundamental states.

\begin{figure}[h]
	\includegraphics[scale=1.15]{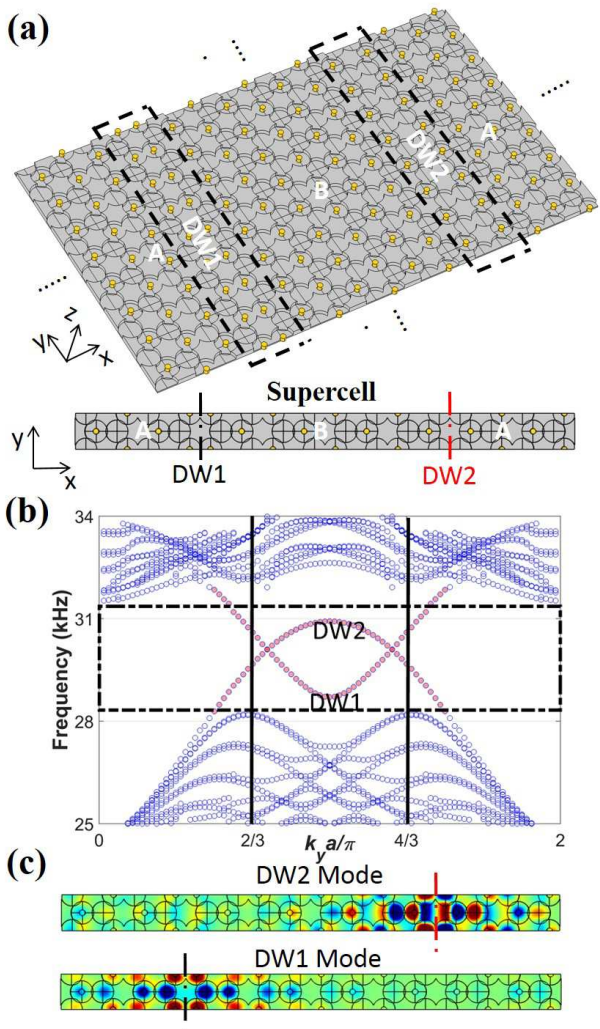}
	\caption{(a) Schematics of an elastic waveguide having $A$- and $B$-state lattices connected along their zigzag edges. This configuration gives rise to two different domain walls DW1 and DW2, as marked by the dashed black boxes. The primitive supercell of such waveguide is also shown. (b) The dispersion relations of the waveguide clearly show the existence of edge states in the topological band. For clarity, only the flexural modes are drawn in the dispersion plot. (c) Plots of the eigenstates of the edge modes supported at the domain walls DW1 and DW2. The plots illustrate different symmetry patterns (either symmetric or antisymmetric) taking place with respect to the interface plane (marked by dashed lines).}\label{Fig4}
\end{figure}

Figure \ref{Fig4}a shows the schematic of an elastic waveguide assembled by connecting $A$- and $B$-state lattices along their zigzag edge. Two different types of domain walls are formed at the interface between the two states, as marked by the black dashed box. These two types of DW have different geometric configurations depending on the position of the $A$- and $B$-states with respect to the interface. For clarity, these two configurations are labeled DW1 ($B$-state is on the right of the $A$-state) and DW2 ($A$-state is on the right of the $B$-state). The schematic of the primitive supercell of the composite waveguide is provided in Fig. \ref{Fig4}a. This supercell was used in the numerical calculations in order to obtain the dispersion properties of the composite waveguide. More specifically, periodic boundary conditions were applied on the four sides of the supercell before solving for the system eigenvalues. Figure \ref{Fig4}b shows the dispersion relations of such waveguide in a frequency range around the topological bandgap (only the $A_0$ modes are plotted).

The appearance of the edge states at DW1 and DW2 can be easily identified in the dispersion results (Figure \ref{Fig4}b) where the bulk modes are marked in blue and the edge modes in red. The solid black vertical lines mark the projections of the valley points along the direction of the interface ($\boldsymbol{K'}$ maps to 2/3 and $\boldsymbol{K}$ to 4/3). Note that the forward propagating mode supported by DW1 (DW2) emanates from the $\boldsymbol{K}$ ($\boldsymbol{K'}$) valley while the backward propagating mode supported by DW1 (DW2) emanates from the $\boldsymbol{K'}$ ($\boldsymbol{K}$) valley. Such large separation in momentum space between the forward and backward traveling modes enables edge states that are almost completely immune from back-scattering, at least in presence of long range disorder. It is also worth observing that the two DW modes are partially gapped (i.e. they do not cross entirely the topological band) due to the large perturbation produced in the original lattice. Examining the eigenstates associated to these edge modes (Fig.\ref{Fig4}c), we observe that they exhibit different symmetry with respect to the plane of the interface (indicated by the dashed lines in Fig.\ref{Fig4}c). In the following, we will show that this aspect leads to different coupling behavior when the edge states are excited by an external source.

\subsection{Full field numerical simulations}

\begin{figure}[h]
	\includegraphics[scale=0.43]{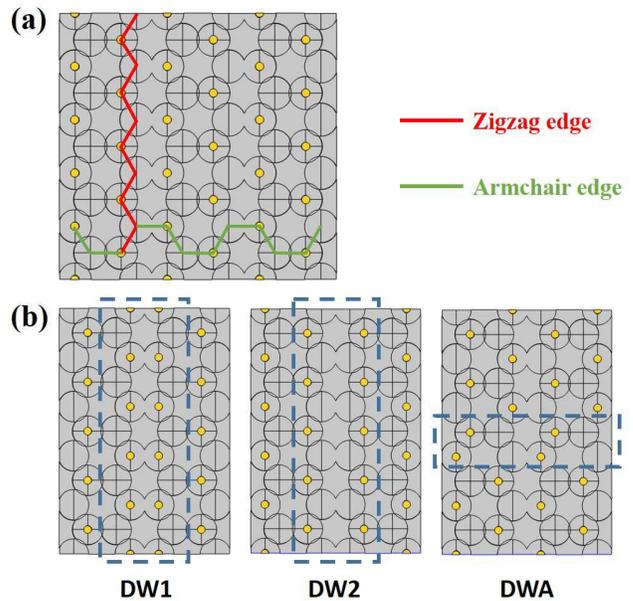}
	\caption{ Schematics illustrating (a) the different edges of the graphene-like lattice and (b) the geometric configurations of different domain walls used to assemble the topological lattice.}\label{FigS1}
\end{figure}

\begin{figure}[h]
	\includegraphics[scale=0.43]{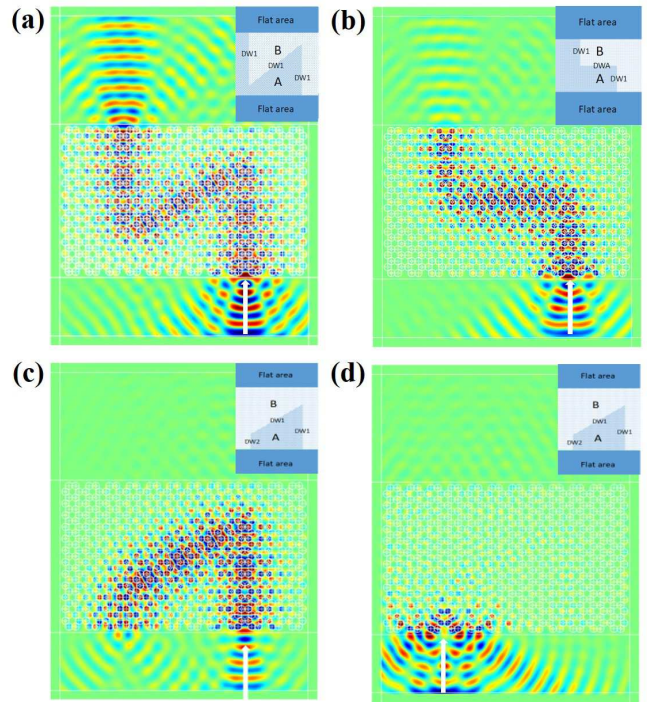}
	\caption{Full field simulations in a flat waveguide integrating an embedded slab of topological material. The system is excited by a plane wave at $f=$29.93 kHz and normal incidence. The insets in each sub-figure provide a schematic view of the different domain wall shapes and configurations, while the white arrow indicates the direction of the incident wave. (a) A Z-shaped domain wall whose individual segments are all DW1 (zigzag type). (b) A Z-shaped domain wall whose segments are a combination of DW1 (zigzag) and DWA (armchair). (c) and (d) show a U-shaped domain wall whose segments are a combination of DW1 and DW2.}\label{Fig5}
\end{figure}

In order to further characterize the propagation behavior of the egde modes, we performed full field numerical simulations on a flat plate having the QVHE topological lattice embedded in the center section. The lattice was excited from a source located in the flat part of the plate (hence external to the topological material) and generating a plane wave at normal incidence. This general configuration was used to investigate the dynamic behavior of different types of DW interfaces. The different interfaces are classified according to the edge-type of the graphene lattice used to assemble the domain wall. In particular, DW1 and DW2 are walls obtained between zigzag edges, and DWA is a wall between armchair edges. Figure \ref{FigS1} provides a schematic view of these domain wall configurations. Numerical results are shown in Fig.\ref{Fig5} for the steady-state response due to an excitation at $f=$29.93 kHz. Perfectly matched layers were applied all around the structure to suppress the boundary reflections. The inset in each sub-figure indicates the domain wall shape being investigated as well as the specific nature of the wall (DW1, DW2, or DWA).

Figure \ref{Fig5}a shows a Z-shape domain wall whose constitutive segments are all made of DW1 type (i.e. zigzag edge) and form two sharp corners having $120^{\circ}$ bends. The numerical results show that the input wave (see white arrow) generates an edge state that is concentrated near the DW and is effectively guided along the wall itself. The transmitted beam appears to have lower intensity compared to the incident beam due to the impedance mismatch between the topological lattice and the flat plate which produces some reflections at both the entrance and exit points. Figure \ref{Fig5}b shows the case of a domain wall characterized by two $90^{\circ}$ bends. This case is different from the former because the mid-segment of the domain wall has armchair edges which theoretically should support a different edge state. This case was explored in order to understand if propagation along dissimilar and concatenated domain walls was possible. Results clearly illustrate the efficient coupling between the zigzag and the armchair edge states, as long as both modes are supported at the same excitation frequency. In Fig.\ref{Fig5}c and d, the domain wall is designed to perform a U-turn. Although, in this case, the individual segments are all of zigzag type, the design mixes DW1 and DW2 along the same interface. This aspect is important because it was previously shown that DW1 and DW2 support edge modes having different symmetry with respect to the interface. 

When the interface is excited on the DW1 interface (Fig.\ref{Fig5}c), which supports symmetric eigenstates, the edge mode can be excited and propagate along the domain wall. On the contrary, when the excitation is applied at DW2 (Fig.\ref{Fig5}d), which supports antisymmetric eigenstates, the incident wave cannot enter the topological interface and it is entirely reflected. This behavior clearly shows the different levels of coupling existing between an external excitation and the different types of domain walls. It also offers a higher flexibility in the design of waveguides having preferential and one-way coupled direction of propagations.

\begin{figure}[h]
	\includegraphics[scale=0.42]{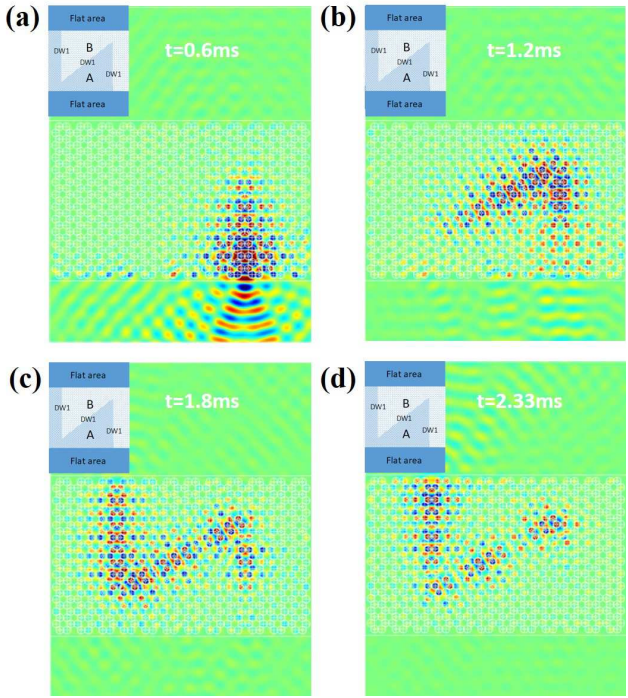}
	\caption{Full field transient simulations showing the propagation of a 20-count wave burst at successive time instants. The conditions for this simulation are identical to those used in Fig. \ref{Fig5}a. (a)-(d) snapshots of the wave field at successive time instants showing that the burst is capable of traveling along sharp bends while giving rise only to very limited back-scattering.}\label{Fig6}
\end{figure}

To further characterize the propagation behavior of the different edge states, we performed a time transient simulation on the same structure used in Fig.\ref{Fig5}a. The main objective was to observe the back-scattering immune behavior of the edge states. A 20-count wave burst having a center frequency of $f=$29.93 kHz was used as excitation. Figure \ref{Fig6} shows snapshots of the propagating wave at successive time instants. Results clearly illustrate how the wave burst is capable of traveling around sharp bends while giving rise only to small amount of back-scattered waves. Note that this result is fully consistent with our previous theoretical analysis that showed a small but nonetheless non-negligible inter-valley mixing between the two counter-propagating edge modes.

\section{Experimental Results}

\begin{figure*}[h!]
	\includegraphics[scale=.8]{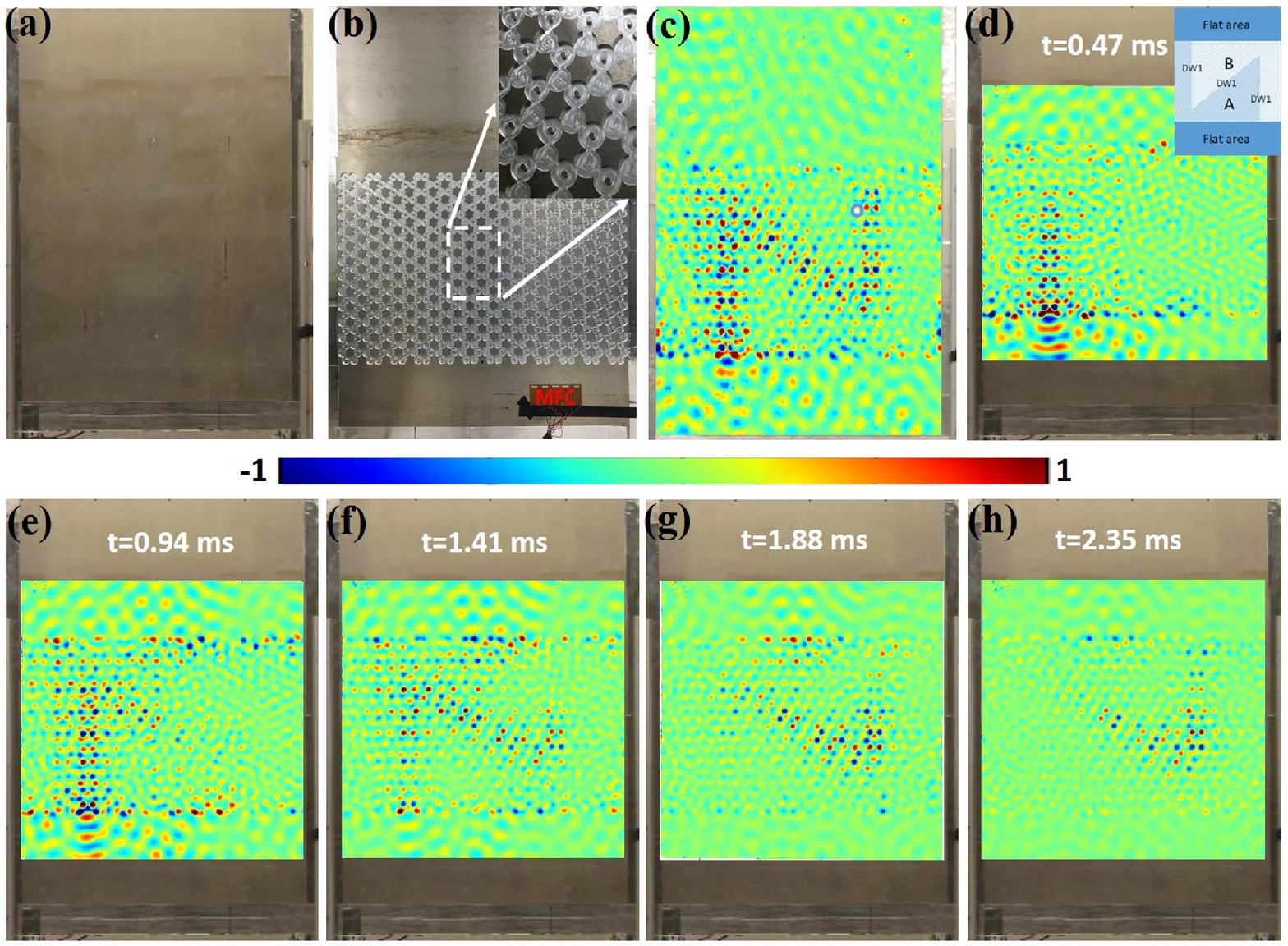}
	\caption{Experimental setup and measurements. (a) Front view of the testbed consisting of a 4.06 mm thick aluminum plate having a slab of QVHE topological material embedded in the center region. (b) shows the back side of the plate sample in order to illustrate the di-atomic graphene-like lattice. An array of MFC patches was surface bonded right in front of the domain wall entrance and used to generate the ultrasonic excitation. The inset provided a detailed zoomed-in view of the lattice structure. (c) Steady state response at an excitation frequency of $f=29.26$ kHz showing the measured transmitted $A_0$ wave field (out-of-plane component) of the test sample. The edge states are well concentrated near the DWs and are guided along the wall itself. (d)-(h) Time transient response to a 20-count burst excitation having a center frequency $f=29.26$ kHz showing the full wave field at successive time instants. The edge state can successfully propagate through the bends while generating only minor reflections.  }\label{Fig7}
\end{figure*}

In order to validate our theoretical and numerical approach to the design of topological waveguides, we performed an experimental validation of one of the designs discussed above. In particular, we selected the Z-shaped interface with segments aligned with the DW1 zigzag edges. The experimental sample was fabricated by CNC machining starting from an initially flat aluminum plate having a thickness of $4.06$mm  (Fig.\ref{Fig7}b). The experimental sample was mounted vertically in an aluminum frame while viscoelastic tape was applied on the surrounding boundaries in order to minimize reflections and reverberation (Fig.\ref{Fig7}a). An array of Micro Fiber Composite (MFC) patches (Fig.\ref{Fig7}b) was surface bonded on the plate in correspondence to the entrance of the zigzag edge and actuated to generate a quasi-$A_0$ plane wave. The number of MFC patches was selected in order to generate a wavefront wide enough to excite the entire Z-shaped channel. The out-of-plane response of the plate was acquired using a Polytec PSV-500 scanning laser vibrometer on the flat side. Both steady-state and transient measurements were performed. The steady-state response was collected following a harmonic excitation at $f=29.2$ kHz which belongs to the topological bandgap. The time transient measurement was obtained in response to a 15-count wave-burst excitation signal having a $f=29.2$ kHz center frequency. In this experimental study, we selected a lower number of counts for the wave burst signal compared with the numerical simulations in order to reduce the length of the burst signal and therefore better visualizing the propagation along the corners and any eventual backscattering.

Figure \ref{Fig7}c shows the experimental measurements under steady-state excitation. The topological edge state along the Z-shaped domain wall is clearly visible accompanied by a very limited penetration of energy into the bulk, as predicted by both theory and simulations. Following the same approach set in place for the numerical simulations, a time transient measurement was also performed in order to observe the possible presence of back-scattering at the sharp corners. Figure \ref{Fig7}d-h show the flexural wave measured at successive time instants. Results clearly indicate that the wave burst is capable of traveling along the Z-shaped path while giving rise only to minor back-scattering. Note that the transient excitation tends to generate a wider spectrum of frequencies in the initial burst. Hence, in order to eliminate the initial broadband transient the measured data were bandpass filtered around the center frequency of the burst.
As a general observation, these results are well consistent with the theoretical and numerical predictions and confirm the possibility to create fully-continuous, load-bearing structural waveguides exhibiting high-level topological properties. \\

\section{Conclusions}

We presented the design and the experimental validation of a fully-continuous and load-bearing thin-walled structural waveguide capable of topologically protected elastic modes. The waveguide was designed by exploiting a concept that is the elastic analogue of the Quantum Valley Hall Effect (QVHE) achieved via a diatomic-graphene-inspired design implemented by using only geometric tapers. Such design preserves the structural properties of the host waveguide while providing a largely increased design space capable of a highly enhanced tailoring of the energy propagation properties.
The diatomic graphene design exhibits a $C_3$ symmetry which breaks the mirror symmetry characteristic of monoatomic graphene $C_{3v}$. The reduced symmetry guarantees the existence of topologically protected edge waves at the interface between domains having different topological charges. Despite the QVHE effect can only give rise to weak topological properties (because time reversal symmetry is still intact), topological edge states having a substantially reduced back-scattering can be achieved. 

This approach enables the design of simple and robust elastic topological waveguides that can be fully embedded in a host structure in order to achieve precise and (quasi) one-directional wave guiding. This approach holds great potential to control the propagation of acoustic waves within structural elements which is a critical capability to enable passive-adaptive vibration and structure-borne noise control in structural systems for high-performance applications.

\section{Acknowledgments}
The authors gratefully acknowledge the financial support of the Air Force Office of Scientific Research under the grant YIP FA9550-15-1-0133.

\bibliography{refbib}% Produces the bibliography via BibTeX.

\begin{thebibliography}{27}%
\makeatletter
\providecommand \@ifxundefined [1]{%
 \@ifx{#1\undefined}
}%
\providecommand \@ifnum [1]{%
 \ifnum #1\expandafter \@firstoftwo
 \else \expandafter \@secondoftwo
 \fi
}%
\providecommand \@ifx [1]{%
 \ifx #1\expandafter \@firstoftwo
 \else \expandafter \@secondoftwo
 \fi
}%
\providecommand \natexlab [1]{#1}%
\providecommand \enquote  [1]{``#1''}%
\providecommand \bibnamefont  [1]{#1}%
\providecommand \bibfnamefont [1]{#1}%
\providecommand \citenamefont [1]{#1}%
\providecommand \href@noop [0]{\@secondoftwo}%
\providecommand \href [0]{\begingroup \@sanitize@url \@href}%
\providecommand \@href[1]{\@@startlink{#1}\@@href}%
\providecommand \@@href[1]{\endgroup#1\@@endlink}%
\providecommand \@sanitize@url [0]{\catcode `\\12\catcode `\$12\catcode
  `\&12\catcode `\#12\catcode `\^12\catcode `\_12\catcode `\%12\relax}%
\providecommand \@@startlink[1]{}%
\providecommand \@@endlink[0]{}%
\providecommand \url  [0]{\begingroup\@sanitize@url \@url }%
\providecommand \@url [1]{\endgroup\@href {#1}{\urlprefix }}%
\providecommand \urlprefix  [0]{URL }%
\providecommand \Eprint [0]{\href }%
\providecommand \doibase [0]{http://dx.doi.org/}%
\providecommand \selectlanguage [0]{\@gobble}%
\providecommand \bibinfo  [0]{\@secondoftwo}%
\providecommand \bibfield  [0]{\@secondoftwo}%
\providecommand \translation [1]{[#1]}%
\providecommand \BibitemOpen [0]{}%
\providecommand \bibitemStop [0]{}%
\providecommand \bibitemNoStop [0]{.\EOS\space}%
\providecommand \EOS [0]{\spacefactor3000\relax}%
\providecommand \BibitemShut  [1]{\csname bibitem#1\endcsname}%
\let\auto@bib@innerbib\@empty
%</preamble>
\bibitem [{\citenamefont {Hasan}\ and\ \citenamefont
  {Kane}(2010)}]{ReviewKaneTI}%
  \BibitemOpen
  \bibfield  {author} {\bibinfo {author} {\bibfnamefont {M.~Z.}\ \bibnamefont
  {Hasan}}\ and\ \bibinfo {author} {\bibfnamefont {C.~L.}\ \bibnamefont
  {Kane}},\ }\href@noop {} {\bibfield  {journal} {\bibinfo  {journal} {Rev.
  Mod. Phys.}\ }\textbf {\bibinfo {volume} {82}},\ \bibinfo {pages} {3045}
  (\bibinfo {year} {2010})}\BibitemShut {NoStop}%
\bibitem [{\citenamefont {Nobelprize.org}(2016)}]{nobel}%
  \BibitemOpen
  \bibfield  {author} {\bibinfo {author} {\bibnamefont {Nobelprize.org}},\
  }\href
  {http://www.nobelprize.org/nobel_prizes/physics/laureates/2016/press.html}
  {\enquote {\bibinfo {title} {The 2016 nobel prize in physics - press
  release},}\ } (\bibinfo {year} {2016})\BibitemShut {NoStop}%
\bibitem [{\citenamefont {Wang}\ \emph {et~al.}(2015)\citenamefont {Wang},
  \citenamefont {Lu},\ and\ \citenamefont {Bertoldi}}]{TopoPhon-Gyro}%
  \BibitemOpen
  \bibfield  {author} {\bibinfo {author} {\bibfnamefont {P.}~\bibnamefont
  {Wang}}, \bibinfo {author} {\bibfnamefont {L.}~\bibnamefont {Lu}}, \ and\
  \bibinfo {author} {\bibfnamefont {K.}~\bibnamefont {Bertoldi}},\ }\href@noop
  {} {\bibfield  {journal} {\bibinfo  {journal} {Phys. Rev. Lett.}\ }\textbf
  {\bibinfo {volume} {115}},\ \bibinfo {pages} {104302} (\bibinfo {year}
  {2015})}\BibitemShut {NoStop}%
\bibitem [{\citenamefont {Nasha}\ \emph {et~al.}(2015)\citenamefont {Nasha},
  \citenamefont {Klecknera}, \citenamefont {Reada}, \citenamefont {Vitellib},
  \citenamefont {Turnerc},\ and\ \citenamefont {Irvine}}]{TopoGyroExp}%
  \BibitemOpen
  \bibfield  {author} {\bibinfo {author} {\bibfnamefont {L.~M.}\ \bibnamefont
  {Nasha}}, \bibinfo {author} {\bibfnamefont {D.}~\bibnamefont {Klecknera}},
  \bibinfo {author} {\bibfnamefont {A.}~\bibnamefont {Reada}}, \bibinfo
  {author} {\bibfnamefont {V.}~\bibnamefont {Vitellib}}, \bibinfo {author}
  {\bibfnamefont {A.~M.}\ \bibnamefont {Turnerc}}, \ and\ \bibinfo {author}
  {\bibfnamefont {W.~T.~M.}\ \bibnamefont {Irvine}},\ }\href@noop {} {\bibfield
   {journal} {\bibinfo  {journal} {Proc. Natl. Acad. Sci. U.S.A.}\ }\textbf
  {\bibinfo {volume} {112}},\ \bibinfo {pages} {14495} (\bibinfo {year}
  {2015})}\BibitemShut {NoStop}%
\bibitem [{\citenamefont {Khanikaev}\ \emph {et~al.}(2015)\citenamefont
  {Khanikaev}, \citenamefont {Fleury}, \citenamefont {Mousavi},\ and\
  \citenamefont {Alu}}]{TopoSound-Flow}%
  \BibitemOpen
  \bibfield  {author} {\bibinfo {author} {\bibfnamefont {A.~B.}\ \bibnamefont
  {Khanikaev}}, \bibinfo {author} {\bibfnamefont {R.}~\bibnamefont {Fleury}},
  \bibinfo {author} {\bibfnamefont {S.~H.}\ \bibnamefont {Mousavi}}, \ and\
  \bibinfo {author} {\bibfnamefont {A.}~\bibnamefont {Alu}},\ }\href@noop {}
  {\bibfield  {journal} {\bibinfo  {journal} {Nat. Comms.}\ }\textbf {\bibinfo
  {volume} {6}},\ \bibinfo {pages} {8260} (\bibinfo {year} {2015})}\BibitemShut
  {NoStop}%
\bibitem [{\citenamefont {Yang}\ \emph {et~al.}(2015)\citenamefont {Yang},
  \citenamefont {Gao}, \citenamefont {Shi}, \citenamefont {Lin}, \citenamefont
  {Gao}, \citenamefont {Chong},\ and\ \citenamefont
  {Zhang}}]{TopologicalAcoustics-Flow}%
  \BibitemOpen
  \bibfield  {author} {\bibinfo {author} {\bibfnamefont {Z.}~\bibnamefont
  {Yang}}, \bibinfo {author} {\bibfnamefont {F.}~\bibnamefont {Gao}}, \bibinfo
  {author} {\bibfnamefont {X.}~\bibnamefont {Shi}}, \bibinfo {author}
  {\bibfnamefont {X.}~\bibnamefont {Lin}}, \bibinfo {author} {\bibfnamefont
  {Z.}~\bibnamefont {Gao}}, \bibinfo {author} {\bibfnamefont {Y.}~\bibnamefont
  {Chong}}, \ and\ \bibinfo {author} {\bibfnamefont {B.}~\bibnamefont
  {Zhang}},\ }\href@noop {} {\bibfield  {journal} {\bibinfo  {journal} {Phys.
  Rev. Lett.}\ }\textbf {\bibinfo {volume} {114}},\ \bibinfo {pages} {114301}
  (\bibinfo {year} {2015})}\BibitemShut {NoStop}%
\bibitem [{\citenamefont {Chen}\ and\ \citenamefont
  {Wu}(2016)}]{TunableTopoPnC-Flow}%
  \BibitemOpen
  \bibfield  {author} {\bibinfo {author} {\bibfnamefont {Z.-G.}\ \bibnamefont
  {Chen}}\ and\ \bibinfo {author} {\bibfnamefont {Y.}~\bibnamefont {Wu}},\
  }\href@noop {} {\bibfield  {journal} {\bibinfo  {journal} {Phys. Rev. Appl.}\
  }\textbf {\bibinfo {volume} {5}},\ \bibinfo {pages} {054021} (\bibinfo {year}
  {2016})}\BibitemShut {NoStop}%
\bibitem [{\citenamefont {S{\"u}sstrunk}\ and\ \citenamefont
  {Huber}(2015)}]{MechanicalTI}%
  \BibitemOpen
  \bibfield  {author} {\bibinfo {author} {\bibfnamefont {R.}~\bibnamefont
  {S{\"u}sstrunk}}\ and\ \bibinfo {author} {\bibfnamefont {S.~D.}\ \bibnamefont
  {Huber}},\ }\href@noop {} {\bibfield  {journal} {\bibinfo  {journal}
  {Science}\ }\textbf {\bibinfo {volume} {349}},\ \bibinfo {pages} {47}
  (\bibinfo {year} {2015})}\BibitemShut {NoStop}%
\bibitem [{\citenamefont {Mousavi}\ \emph {et~al.}(2015)\citenamefont
  {Mousavi}, \citenamefont {Khanikaev},\ and\ \citenamefont
  {Wang}}]{AcousticTIPlate}%
  \BibitemOpen
  \bibfield  {author} {\bibinfo {author} {\bibfnamefont {S.~H.}\ \bibnamefont
  {Mousavi}}, \bibinfo {author} {\bibfnamefont {A.~B.}\ \bibnamefont
  {Khanikaev}}, \ and\ \bibinfo {author} {\bibfnamefont {Z.}~\bibnamefont
  {Wang}},\ }\href@noop {} {\bibfield  {journal} {\bibinfo  {journal} {Nat.
  Commun.}\ }\textbf {\bibinfo {volume} {6}},\ \bibinfo {pages} {8682}
  (\bibinfo {year} {2015})}\BibitemShut {NoStop}%
\bibitem [{\citenamefont {He}\ \emph {et~al.}(2016)\citenamefont {He},
  \citenamefont {Ni}, \citenamefont {Ge}, \citenamefont {Sun}, \citenamefont
  {Chen}, \citenamefont {Lu}, \citenamefont {Liu},\ and\ \citenamefont
  {Chen}}]{AcousticTIAir}%
  \BibitemOpen
  \bibfield  {author} {\bibinfo {author} {\bibfnamefont {C.}~\bibnamefont
  {He}}, \bibinfo {author} {\bibfnamefont {X.}~\bibnamefont {Ni}}, \bibinfo
  {author} {\bibfnamefont {H.}~\bibnamefont {Ge}}, \bibinfo {author}
  {\bibfnamefont {X.-C.}\ \bibnamefont {Sun}}, \bibinfo {author} {\bibfnamefont
  {Y.-B.}\ \bibnamefont {Chen}}, \bibinfo {author} {\bibfnamefont {M.-H.}\
  \bibnamefont {Lu}}, \bibinfo {author} {\bibfnamefont {X.-P.}\ \bibnamefont
  {Liu}}, \ and\ \bibinfo {author} {\bibfnamefont {Y.-F.}\ \bibnamefont
  {Chen}},\ }\href@noop {} {\bibfield  {journal} {\bibinfo  {journal} {Nat.
  Phys.}\ }\textbf {\bibinfo {volume} {12}},\ \bibinfo {pages} {1124} (\bibinfo
  {year} {2016})}\BibitemShut {NoStop}%
\bibitem [{\citenamefont {Lu}\ \emph {et~al.}(2016{\natexlab{a}})\citenamefont
  {Lu}, \citenamefont {Qiu}, \citenamefont {Ke},\ and\ \citenamefont
  {Liu}}]{ValleySonicBulk}%
  \BibitemOpen
  \bibfield  {author} {\bibinfo {author} {\bibfnamefont {J.}~\bibnamefont
  {Lu}}, \bibinfo {author} {\bibfnamefont {C.}~\bibnamefont {Qiu}}, \bibinfo
  {author} {\bibfnamefont {M.}~\bibnamefont {Ke}}, \ and\ \bibinfo {author}
  {\bibfnamefont {Z.}~\bibnamefont {Liu}},\ }\href@noop {} {\bibfield
  {journal} {\bibinfo  {journal} {Phys. Rev. Lett.}\ }\textbf {\bibinfo
  {volume} {116}},\ \bibinfo {pages} {093901} (\bibinfo {year}
  {2016}{\natexlab{a}})}\BibitemShut {NoStop}%
\bibitem [{\citenamefont {Lu}\ \emph {et~al.}(2016{\natexlab{b}})\citenamefont
  {Lu}, \citenamefont {Qiu}, \citenamefont {Ye}, \citenamefont {Fan},
  \citenamefont {Ke}, \citenamefont {Zhang},\ and\ \citenamefont
  {Liu}}]{ValleySonicEdge}%
  \BibitemOpen
  \bibfield  {author} {\bibinfo {author} {\bibfnamefont {J.}~\bibnamefont
  {Lu}}, \bibinfo {author} {\bibfnamefont {C.}~\bibnamefont {Qiu}}, \bibinfo
  {author} {\bibfnamefont {L.}~\bibnamefont {Ye}}, \bibinfo {author}
  {\bibfnamefont {X.}~\bibnamefont {Fan}}, \bibinfo {author} {\bibfnamefont
  {M.}~\bibnamefont {Ke}}, \bibinfo {author} {\bibfnamefont {F.}~\bibnamefont
  {Zhang}}, \ and\ \bibinfo {author} {\bibfnamefont {Z.}~\bibnamefont {Liu}},\
  }\href@noop {} {\bibfield  {journal} {\bibinfo  {journal} {Nat. Phys.}\
  }\textbf {\bibinfo {volume} {13}},\ \bibinfo {pages} {369} (\bibinfo {year}
  {2016}{\natexlab{b}})}\BibitemShut {NoStop}%
\bibitem [{\citenamefont {Pal}\ and\ \citenamefont {Ruzzene}(2017)}]{Ruzz1}%
  \BibitemOpen
  \bibfield  {author} {\bibinfo {author} {\bibfnamefont {R.~K.}\ \bibnamefont
  {Pal}}\ and\ \bibinfo {author} {\bibfnamefont {M.}~\bibnamefont {Ruzzene}},\
  }\href@noop {} {\bibfield  {journal} {\bibinfo  {journal} {New Journal of
  Physics}\ }\textbf {\bibinfo {volume} {19}},\ \bibinfo {pages} {025001}
  (\bibinfo {year} {2017})}\BibitemShut {NoStop}%
\bibitem [{\citenamefont {Vila}\ \emph {et~al.}(2017)\citenamefont {Vila},
  \citenamefont {Pal},\ and\ \citenamefont {Ruzzene}}]{Ruzz2}%
  \BibitemOpen
  \bibfield  {author} {\bibinfo {author} {\bibfnamefont {J.}~\bibnamefont
  {Vila}}, \bibinfo {author} {\bibfnamefont {R.~K.}\ \bibnamefont {Pal}}, \
  and\ \bibinfo {author} {\bibfnamefont {M.}~\bibnamefont {Ruzzene}},\
  }\href@noop {} {\bibfield  {journal} {\bibinfo  {journal} {Phys. Rev. B}\
  }\textbf {\bibinfo {volume} {96}},\ \bibinfo {pages} {134307} (\bibinfo
  {year} {2017})}\BibitemShut {NoStop}%
\bibitem [{\citenamefont {Liu}\ and\ \citenamefont
  {Semperlotti}(2017)}]{TWarXiv}%
  \BibitemOpen
  \bibfield  {author} {\bibinfo {author} {\bibfnamefont {T.-W.}\ \bibnamefont
  {Liu}}\ and\ \bibinfo {author} {\bibfnamefont {F.}~\bibnamefont
  {Semperlotti}},\ }\href@noop {} {\bibfield  {journal} {\bibinfo  {journal}
  {arXiv preprint arXiv:1708.02987}\ } (\bibinfo {year} {2017})}\BibitemShut
  {NoStop}%
\bibitem [{\citenamefont {Raghu}\ and\ \citenamefont {Haldane}(2008)}]{Raghu}%
  \BibitemOpen
  \bibfield  {author} {\bibinfo {author} {\bibfnamefont {S.}~\bibnamefont
  {Raghu}}\ and\ \bibinfo {author} {\bibfnamefont {F.~D.~M.}\ \bibnamefont
  {Haldane}},\ }\href@noop {} {\bibfield  {journal} {\bibinfo  {journal} {Phys.
  Rev. A}\ }\textbf {\bibinfo {volume} {78}},\ \bibinfo {pages} {033834}
  (\bibinfo {year} {2008})}\BibitemShut {NoStop}%
\bibitem [{\citenamefont {Ochiai}\ and\ \citenamefont
  {Onoda}(2009)}]{PhotonicGraphene}%
  \BibitemOpen
  \bibfield  {author} {\bibinfo {author} {\bibfnamefont {T.}~\bibnamefont
  {Ochiai}}\ and\ \bibinfo {author} {\bibfnamefont {M.}~\bibnamefont {Onoda}},\
  }\href@noop {} {\bibfield  {journal} {\bibinfo  {journal} {Phys. Rev. B}\
  }\textbf {\bibinfo {volume} {80}},\ \bibinfo {pages} {155103} (\bibinfo
  {year} {2009})}\BibitemShut {NoStop}%
\bibitem [{\citenamefont {Kane}\ and\ \citenamefont {Mele}(2005)}]{Z2}%
  \BibitemOpen
  \bibfield  {author} {\bibinfo {author} {\bibfnamefont {C.~L.}\ \bibnamefont
  {Kane}}\ and\ \bibinfo {author} {\bibfnamefont {E.~J.}\ \bibnamefont
  {Mele}},\ }\href@noop {} {\bibfield  {journal} {\bibinfo  {journal} {Phys.
  Rev. Lett.}\ }\textbf {\bibinfo {volume} {95}},\ \bibinfo {pages} {146802}
  (\bibinfo {year} {2005})}\BibitemShut {NoStop}%
\bibitem [{\citenamefont {Sheng}\ \emph {et~al.}(2006)\citenamefont {Sheng},
  \citenamefont {Weng}, \citenamefont {Sheng},\ and\ \citenamefont
  {Haldane}}]{SpinChernHaldane}%
  \BibitemOpen
  \bibfield  {author} {\bibinfo {author} {\bibfnamefont {D.~N.}\ \bibnamefont
  {Sheng}}, \bibinfo {author} {\bibfnamefont {Z.}~\bibnamefont {Weng}},
  \bibinfo {author} {\bibfnamefont {L.}~\bibnamefont {Sheng}}, \ and\ \bibinfo
  {author} {\bibfnamefont {F.~D.~M.}\ \bibnamefont {Haldane}},\ }\href@noop {}
  {\bibfield  {journal} {\bibinfo  {journal} {Phys. Rev. Lett.}\ }\textbf
  {\bibinfo {volume} {97}},\ \bibinfo {pages} {036808} (\bibinfo {year}
  {2006})}\BibitemShut {NoStop}%
\bibitem [{\citenamefont {Zhu}\ and\ \citenamefont {Semperlotti}(2013)}]{Zhu}%
  \BibitemOpen
  \bibfield  {author} {\bibinfo {author} {\bibfnamefont {H.}~\bibnamefont
  {Zhu}}\ and\ \bibinfo {author} {\bibfnamefont {F.}~\bibnamefont
  {Semperlotti}},\ }\href@noop {} {\bibfield  {journal} {\bibinfo  {journal}
  {AIP Advances}\ }\textbf {\bibinfo {volume} {3}},\ \bibinfo {pages} {092121}
  (\bibinfo {year} {2013})}\BibitemShut {NoStop}%
\bibitem [{\citenamefont {Semperlotti}\ and\ \citenamefont
  {Zhu}(2014)}]{Semperlotti}%
  \BibitemOpen
  \bibfield  {author} {\bibinfo {author} {\bibfnamefont {F.}~\bibnamefont
  {Semperlotti}}\ and\ \bibinfo {author} {\bibfnamefont {H.}~\bibnamefont
  {Zhu}},\ }\href@noop {} {\bibfield  {journal} {\bibinfo  {journal} {Journal
  of Applied Physics}\ }\textbf {\bibinfo {volume} {116}},\ \bibinfo {pages}
  {054906} (\bibinfo {year} {2014})}\BibitemShut {NoStop}%
\bibitem [{\citenamefont {Zhu}\ and\ \citenamefont {Semperlotti}(2014)}]{Zhu2}%
  \BibitemOpen
  \bibfield  {author} {\bibinfo {author} {\bibfnamefont {H.}~\bibnamefont
  {Zhu}}\ and\ \bibinfo {author} {\bibfnamefont {F.}~\bibnamefont
  {Semperlotti}},\ }\href@noop {} {\bibfield  {journal} {\bibinfo  {journal}
  {Journal of Applied Physics}\ }\textbf {\bibinfo {volume} {116}},\ \bibinfo
  {pages} {094901} (\bibinfo {year} {2014})}\BibitemShut {NoStop}%
\bibitem [{\citenamefont {Zhu}\ and\ \citenamefont
  {Semperlotti}(2015)}]{ZhuPRB}%
  \BibitemOpen
  \bibfield  {author} {\bibinfo {author} {\bibfnamefont {H.}~\bibnamefont
  {Zhu}}\ and\ \bibinfo {author} {\bibfnamefont {F.}~\bibnamefont
  {Semperlotti}},\ }\href@noop {} {\bibfield  {journal} {\bibinfo  {journal}
  {Phys. Rev. B}\ }\textbf {\bibinfo {volume} {91}},\ \bibinfo {pages} {104304}
  (\bibinfo {year} {2015})}\BibitemShut {NoStop}%
\bibitem [{\citenamefont {Zhu}\ and\ \citenamefont
  {Semperlotti}(2016)}]{ZhuPRL}%
  \BibitemOpen
  \bibfield  {author} {\bibinfo {author} {\bibfnamefont {H.}~\bibnamefont
  {Zhu}}\ and\ \bibinfo {author} {\bibfnamefont {F.}~\bibnamefont
  {Semperlotti}},\ }\href@noop {} {\bibfield  {journal} {\bibinfo  {journal}
  {Phys. Rev. Lett.}\ }\textbf {\bibinfo {volume} {117}},\ \bibinfo {pages}
  {034302} (\bibinfo {year} {2016})}\BibitemShut {NoStop}%
\bibitem [{\citenamefont {Zhu}\ and\ \citenamefont
  {Semperlotti}(2017)}]{ZhuJAP}%
  \BibitemOpen
  \bibfield  {author} {\bibinfo {author} {\bibfnamefont {H.}~\bibnamefont
  {Zhu}}\ and\ \bibinfo {author} {\bibfnamefont {F.}~\bibnamefont
  {Semperlotti}},\ }\href@noop {} {\bibfield  {journal} {\bibinfo  {journal}
  {arXiv preprint arXiv:1701.03445}\ } (\bibinfo {year} {2017})}\BibitemShut
  {NoStop}%
\bibitem [{\citenamefont {Xiao}\ \emph {et~al.}(2007)\citenamefont {Xiao},
  \citenamefont {Yao},\ and\ \citenamefont {Niu}}]{ValleyContrasting}%
  \BibitemOpen
  \bibfield  {author} {\bibinfo {author} {\bibfnamefont {D.}~\bibnamefont
  {Xiao}}, \bibinfo {author} {\bibfnamefont {W.}~\bibnamefont {Yao}}, \ and\
  \bibinfo {author} {\bibfnamefont {Q.}~\bibnamefont {Niu}},\ }\href@noop {}
  {\bibfield  {journal} {\bibinfo  {journal} {Phys. Rev. Lett.}\ }\textbf
  {\bibinfo {volume} {99}},\ \bibinfo {pages} {236809} (\bibinfo {year}
  {2007})}\BibitemShut {NoStop}%
\bibitem [{\citenamefont {Yao}\ \emph {et~al.}(2009)\citenamefont {Yao},
  \citenamefont {Yang},\ and\ \citenamefont {Niu}}]{ControlEdgeStates}%
  \BibitemOpen
  \bibfield  {author} {\bibinfo {author} {\bibfnamefont {W.}~\bibnamefont
  {Yao}}, \bibinfo {author} {\bibfnamefont {S.~A.}\ \bibnamefont {Yang}}, \
  and\ \bibinfo {author} {\bibfnamefont {Q.}~\bibnamefont {Niu}},\ }\href@noop
  {} {\bibfield  {journal} {\bibinfo  {journal} {Phys. Rev. Lett.}\ }\textbf
  {\bibinfo {volume} {102}},\ \bibinfo {pages} {096801} (\bibinfo {year}
  {2009})}\BibitemShut {NoStop}%
\end{thebibliography}%

\end{document}